\documentclass{aa}
\usepackage{graphicx}
\usepackage{txfonts}
\usepackage{natbib}
\usepackage{multirow}
\usepackage{hyperref} 
\usepackage{mathrsfs}
\hypersetup{colorlinks=true,linkcolor=blue,citecolor=blue,urlcolor=blue}
\usepackage{orcidlink}

\begin{document}

\title{The effects of yields from binary massive stars as functions of metallicity}

\author {E. Pepe \orcidlink{0009-0006-9760-3335} \inst{1} \thanks{email to: EMANUELE.PEPE@phd.units.it},  
E. Spitoni \orcidlink{0000-0001-9715-5727}\inst{2, 3},  
F. Matteucci \orcidlink{0000-0001-7067-2302} \inst{1,2,4} \and
M. Palla \orcidlink{0000-0002-3574-9578}\inst{5}}

\institute{
   Dipartimento di Fisica, Sezione di Astronomia,
  Universit\`a di Trieste, Via G.~B. Tiepolo 11, I-34143 Trieste   \and 
  INAF - Osservatorio Astronomico di Trieste, via G.B. Tiepolo
 11, I-34131, Trieste, Italy
 \and 
 IFPU Institute for Fundamental Physics of the Universe, Via Beirut 2, I-34151 Trieste, Italy
 \and 
 INFN - Sezione di Trieste, via Valerio 2, I-34134 Trieste, Italy
 \and 
 INAF - Osservatorio Astrofisico di Arcetri, Largo E. Fermi 5, 50125, Firenze, Italy\\
}

  \date{Received xxxx / Accepted xxxx}

\abstract {Massive stars in binary systems that undergo mass transfer during  their lifetime have a different evolution from that of single stars, possibly affecting their chemical yields. While massive stars produce most of the metals in the Universe, only few studies have investigated the effects of massive binary stars on the chemical evolution of the Milky Way. Following the most recent studies on massive binary-stripped star yields as functions of metallicity, we aim at improving previous results based on single-metallicity model grids. Here, by adopting a detailed model of chemical evolution for our Galaxy, we compute the evolution of 22 chemical species including C, N, O, $\alpha$-elements and Fe-peak elements, adopting novel prescriptions for  single and binary massive star yields. 
Our main results can be summarised as follows: (i) consistently with previous predictions, we observe very small differences in both the predicted solar abundances and [X/Fe] vs [Fe/H] relations  even when including massive binary yields depending on metallicity;  (ii)  when adopting the new set of stellar yields for massive single stars, as computed by \citet{farmer2023}, we  are able to reproduce both the K solar abundance as well as the [K/Fe] vs [Fe/H] relation, without invoking ad hoc assumptions on nucleosynthesis prescriptions; (iii) our model adopting Farmer's yields both for single and binary massive stars is able to better reproduce the [X/Fe] versus [Fe/H] relation for both Mg and Ca, as compared with standard nucleosynthetic yields adopted in chemical evolution models; iv) we find that no models can well reproduce the [C/Fe] and [Ti/Fe] vs [Fe/H] when adopting the new yields as functions of metallicity.}

\keywords{Galaxy: disc -- Galaxy: evolution  -- Nuclear reactions, nucleosynthesis, abundances -- Galaxy: abundances}

\titlerunning{Yields from massive binaries}

\authorrunning{E. Pepe et al.}

\maketitle
\section{Introduction}
The aim of Galactic archaeology is to reconstruct the star formation history and chemical evolution of the Milky Way (MW) and external galaxies mainly through the study of chemical abundances observed in stars and in the interstellar medium (ISM). A key ingredient is the amount of chemical elements produced by stars and subsequently released into the ISM, namely the stellar yields. While the stellar yields are fundamental to any model of galactic chemical evolution, they also represent one of the main sources of intrinsic uncertainty (see, e.g., \citealt{romano2010, matteucci2021,palla2021}, and references therein).

Massive stars ($M \gtrsim 10\,M_\odot$) are responsible for the production of most of the metals in galaxies, which they eject into the ISM through stellar winds and  mainly by core-collapse supernovae (CC-SNe). Although detailed chemical evolution models track the synthesis of a wide range of elements, from hydrogen up to $^{63}$Cu and $^{64}$Zn, they generally rely on stellar yields computed for single massive stars (e.g., \citealt{WW1995, koba2006, nomoto2013, limongi2018}; see however \citealt{dedonder&vanbeveren2002,dedonder2004}). This approximation is mainly driven by the limited availability of extensive grids of yields for binary stellar models, despite the fact that stars predominantly form in clustered environments \citep{lada2003} and that binary systems are very common. Binary interactions can significantly alter the stellar evolutionary paths and nucleosynthetic outputs relative to single-star evolution, through multiple interaction phases such as mass transfer, envelope stripping, and common-envelope evolution (see, e.g., \citealt{dedonder&vanbeveren2002,dedonder2004, langer2012, woosley2019, farmer2021,farmer2023,ma2025}).

To date, only a limited number of studies have investigated the impact of massive binaries on the chemical evolution of the Milky Way, most notably \citet{dedonder&vanbeveren2002,dedonder2004} and \citet{pepe2025}. The former computed chemical yields from massive binary systems and tested them within a two-infall model for the MW similar to that of \citet{chiappini1997}. They found that the inclusion of massive binaries improves the agreement with the observed temporal evolution of carbon and suggested that binary evolution could contribute to the production of primary nitrogen from massive stars, although nitrogen can also be synthesised by single rotating massive stars (e.g., \citealt{meynet2002, limongi2018}). At the same time, \citet{dedonder&vanbeveren2002,dedonder2004} concluded that the inclusion of massive binaries leads to variations of no more than a factor of two in the evolution of other elements (He, O, Ne, Mg, Si, S, and Ca) compared to models without binaries.

More recently, \citet{pepe2025}, from here onward referred to as Paper~I, explored the impact of massive binary-stripped star yields computed by \citet{farmer2023} for  solar metallicity, testing them within both one-infall \citep{chiosi1980, matteucci1986, matteucci1989, boissier1999} and two-infall chemical evolution models for the MW \citep{chiappini1997, spitoni2019}. Paper~I followed the evolution of several elements from He to Zn and found that binarity introduces only modest changes in abundance ratios for most elements, while the dominant differences arise from the choice of massive single-star yields. Notably, their study was the first to successfully reproduce the observed potassium solar abundance and [K/Fe] versus [Fe/H]\footnote{[X/Y] $= \log(X/Y) - \log(X_\odot/Y_\odot)$, where X and Y are the abundances in the object considered and $X_\odot$ and $Y_\odot$ are the corresponding solar abundances.} abundance ratio without invoking ad hoc assumptions, when adopting \citet{farmer2023} yields for single massive stars only. However, it is worth reminding that a major limitation of their analysis was that the \citet{farmer2023} yields were computed only at solar metallicity.

Recently, \citet{ma2025} extended the work of \citet{farmer2021} by computing yields for carbon from massive binary-stripped stars as functions of metallicity, and by proposing an interpolation prescription aimed at describing the expected metallicity dependence of massive binary yields for other elements.

In this paper, we adopt the prescriptions introduced by \citet{ma2025} to extend the analysis of Paper~I and to overcome one of its main limitations. We follow the evolution of several chemical species (H, He, C, N, $\alpha$-elements, Fe, and Fe-peak elements) within a detailed and well-tested chemical evolution model for the Milky Way, which is a revised version of the original model by \citet[][e.g., \citealt{spitoni2019, spitoni2020, palla2020, palla2021}; Paper~I]{chiappini1997}. In particular, we adopt for the first time the metallicity-dependent carbon yields computed by \citet{ma2025} for both single and massive binary-stripped stars, as well as the yields for all other elements from \citet{farmer2023}, combined with the metallicity-dependent prescriptions as suggested by \citet{ma2025}. 

Following Paper~I, we compute the expected solar chemical abundances and the evolution of [X/Fe] versus [Fe/H], where X denotes all elements except Fe, which is commonly used as a tracer of stellar metallicity. The model predictions are compared with the abundance patterns derived from large spectroscopic surveys of the MW disc, in particular APOGEE \citep{apogeedr172022}, as well as smaller, high-resolution surveys \citep{hinkel2014,nissen2020}, in order to assess whether the new prescriptions for the adopted binary yields  lead to an improved agreement between theoretical predictions and observations.

We organise the paper as follows: in section~\ref{mod_sec} we describe the adopted chemical evolution model, with a focus on the new prescriptions for massive binary stripped stellar yields. In section~\ref{data_sec} we describe the observational data adopted throughout this work. In section~\ref{res_sec} we show and discuss the results for our predictions compared to observations, and in section~\ref{conc_sec} we outline our conclusions and final remarks.

\section{Chemical Evolution Model}
\label{mod_sec}
In this section we present the model adopted in this paper. Following the work of Paper~I, we adopt a revised two-infall model \citep[e.g.,][]{palla2020} that assumes that two sequential infall episodes formed the thick and thin disc respectively. The first episode forms the thick disc on a very short timescale of $\tau\sim$1 Gyr, while the second episode forms the thin disc on a longer timescale of $\tau\sim$7 Gyr. We adopt a delay between the two infall episodes of 3.25 Gyr \citep{palla2020}, larger than the 'classical' 1 Gyr  obtained with the two-infall model \citep{chiappini1997, romano2010}.\\

The basic equation we use to describe the evolution of an element $i$ in the ISM is \citep[see][]{matteucci2021}:

 \begin{equation}
    \label{}
    \dot{\sigma}_i(R,t)=-\psi(R,t)\,X_i(R,t)+\dot{ \mathscr{R}}_i(R,t)+\dot{\sigma}_{i,inf}(R,t),
\end{equation}
where on the left-hand side is the fractional surface mass density of the element $i$ in the ISM at the time t, $\sigma_i(R,t)=\sigma_{gas}(R,t)\,X_i(R,t)$ with $X_i(R,t)$ being the surface mass abundance of the given element and $\sigma_{gas}(R,t)$ being the mass density of the ISM. On the right-hand side, the first term is the rate at which the element $i$ is trapped into stars at each epoch, with $\psi(R,t)$ being the star-formation rate (SFR), parametrised according to the Schmidt-Kennicutt law \citep{kenni1998}:

\begin{equation}
\label{}
    \psi(R,t)=\nu(R)\, \sigma_{gas}^k(R,t),
\end{equation}
with $k=1.5$, observationally derived, while $\nu$ is the star formation efficiency expressed in units of Gyr$^{-1}$, variable with the Galactocentric distance and equal to 2 Gyr$^{-1}$ for the solar neighbourhood during the first infall event and to 1 Gyr$^{-1}$ during the second infall \citep[see][]{palla2020}.

The second term of the right side, $\dot{\mathscr{R}}_i(R,t)$, is the rate of restitution of elements from stars into the ISM in the form of new and old element $i$, namely the rate at which chemical elements are returned from stars into the ISM through stellar winds and supernova explosions. $\mathscr{R}_i(R,t)$ also depends on the initial mass function (IMF), here parametrised as in \citet{kroupa1993}.

The last term is the gas infall rate, computed as:

\begin{equation}
    \dot{\sigma}_{i,inf}(R,t)=A(R)\, X_{i,\,inf}\, e^{-\frac{t}{\tau_1}} + \theta(t-t_{max})\, B(R)\, X_{i,\,inf}\, e^{-\frac{t-t_{max}}{\tau_2}}
\end{equation}

where $\tau_1$ and $\tau_2$ are the infall timescale for the first and second episode, respectively. $X_{i,\,inf}$ is the chemical composition of the infalling gas (here assumed to be primordial), while $A(R)$ and $B(R)$ are the coefficients obtained by reproducing the present-day surface mass density of the thick and thin disc in the solar neighbourhood of $47.1 \pm 3.4$ M$_\odot\,$pc$^{-2}$ as proposed by \citet{mckee2015}. Lastly $t_{max}$ is the time of maximum infall onto the thin disc, which is also the delay between the two infall episodes.

Following the same prescriptions of Paper~I, the model includes a detailed computation of the CC-SN rate, as well as Type Ia SN (SNe\,Ia) rate assuming the single degenerate scenario expressed through a delay-time-distribution as described by \citet{MR01}. We also adopt no Galactic winds,  since many studies \citep{melioli2008,melioli2009,spitoni2008,spitoni2009,hopkins2023} found that the metals ejected through Type II Supernova explosions in OB associations within massive disc galaxies,  very probably fall back onto the same Galactocentric region from which they originated, thus rendering Galactic wind ineffective. The gas falling back onto the disc is known as "Galactic fountain" and its effects on the chemical evolution has been studied by \cite{spitoni2008}, who concluded that these fountains do not appreciably affect the chemical evolution of the disc.

\subsection{Stellar yields}
\label{subsec_yields}
This work follows the study of Paper~I which adopted for the first time the stellar yields by \citet{farmer2023} for massive binary stars. 
\citet{farmer2023} estimated stellar yields for both single and binary massive stars for elements up to Zn, for and extensive grid of masses ($M_{init}$= 11 to 45 $M_\odot$), using the MESA stellar evolution code (version 12115, see e.g. \citealt[\citeyear{paxton2013}, \citeyear{paxton2015}, \citeyear{paxton2018}, \citeyear{paxton2019}]{paxton2011}, \citealt{jermyn2023}). \\

In \citet{farmer2023}, stellar models are evolved from the zero-age main sequence up to core collapse and through the ensuing supernova phase until shock breakout. They assume a solar chemical composition following \citet{Grevesse98}. Both single and binary stellar models are computed without rotation.

For binary systems, \citet{farmer2023} model the evolution of the primary star in the presence of a companion with a fixed mass ratio $M_2/M_1 = 0.8$ and initial orbital periods ranging from 38 to 300 days. This choice ensures that all systems experience Case~B mass transfer, namely Roche-lobe overflow occurring after core-hydrogen exhaustion of the primary star (\citealt{paczynski1967}; \citealt{vandenheuvel1969}). During the evolution, the secondary star is treated as a point mass until the end of core-helium burning, after which it is removed from the system and the primary star is evolved in isolation until core collapse (see \citealt{laplace2020}). As a consequence, the resulting nucleosynthetic yields correspond exclusively to the primary star in the binary system. 

Here, we also adopt the $^{12}$C yields as recently computed by \citet{ma2025} as functions of metallicity, referred to as Ma25 hereafter. In their study, they also suggest that yields from binary-stripped stars for other elements can be considered as a function of metallicity by adopting the following formula:

\begin{equation}
\label{maeq}
m_{binary}(Z) = m_{F23} + (m_{F23}-m_{single})\,max\left[1.25\,\log\left(\frac{Z}{Z_\odot}\right),-1\right],
\end{equation}

where $m_{binary}(Z)$ is the massive binary yield as a function of metallicity, $m_{F23}$ is the binary massive star yield as computed in \citet{farmer2023} for the solar chemical composition, and $m_{single}$ is the massive single star yield computed in \citet{farmer2023}.  While this is only an approximation, it represents an improvement on adopting solar metallicity yields for all metallicities.
We will refer to \citet{farmer2023} yields for both single massive and binary stars, modified adopting the prescriptions of eq.~\ref{maeq}, as FM23.

We consider only two fractions of 70\% and 0\% for massive binaries in the IMF: this is because Paper~I found that once the binaries are considered, changing their fraction causes minimal effects on the evolution of the chemical abundances in the ISM. In particular, they tested 0\%, 50\%, 70\% and 100\% binary systems in the IMF of massive stars.\\

Throughout this work, we also adopt yields as suggested in \citet[their model 15]{romano2010} (R10 hereafter) for single massive stars to allow us to make a comparison between the newly proposed yields and well-tested yields for single massive star from the literature. R10 yields consist in a combination of models obtained with the Geneva stellar evolutionary code (\citealt{meynet2002}; \citealt[\citeyear{hirschi2007}]{hirschi2005}; \citealt{ekstrom2008}) for CNO elements, and those of \citet{koba2006} for other elements (see \citealt{romano2010} for more details).

While our focus is on the effect of massive star yields, the model also includes yields for low and intermediate mass stars (LIMS) and SNe\,Ia to properly account for the chemical evolution of the Galaxy. We adopt for all models the yields of \citet{karakas2010} for LIMS and those of \citet[their model W7]{iwamoto1999} for SNe\,Ia.

\section{Observational Data}
\label{data_sec}

In this work we adopt abundances for the solar vicinity from APOGEE DR17 \citep{apogeedr172022}, as well as from smaller, high resolution surveys such as \citet{hinkel2014} and \citet{nissen2020}. In this section, we provide additional details on the different datasets adopted and selected chemical elements to perform our comparison.

\subsection{The APOGEE DR17 data sample}
\label{subsec_apogee}

Throughout this work, we make use of stellar abundance data from the APOGEE DR17 survey \citep{apogeedr172022}, part of the Sloan Digital Sky Survey (SDSS). APOGEE observations are carried out using both the du Pont Telescope and the Sloan Foundation 2.5 m Telescope \citep{gunn2006} at Apache Point Observatory. Stellar atmospheric parameters and chemical abundances are derived with the APOGEE Stellar Parameters and Chemical Abundances Pipeline (ASPCAP; \citealt{gperez2016}). The model atmospheres adopted in DR17 are based on the MARCS models \citep{gustafsson2008}, as described in \citet{jonsson2020}, while the employed atomic and molecular line list is presented in \citet{smith2021}.

We restrict our analysis to stars located in the solar neighbourhood, selecting objects with Galactocentric distances in the range 7\,kpc $\le R_{\rm GC} \le 9$\,kpc, as computed by \citet{leung2019} and provided in the astroNN value-added catalogue\footnote{\url{https://data.sdss.org/sas/dr17/apogee/vac/apogee-astronn/}}. In this catalogue, stellar distances, particularly for more distant objects, are estimated using a deep neural network trained on Gaia parallaxes for stars in common between Gaia (\citealt{gaia2016,Egaia2021}) and APOGEE. Following \citet{spitoni2024} and Paper~I, we further apply quality cuts based on signal-to-noise ratio and vertical distance from the Galactic plane, retaining only stars with SNR $> 80$ and $|z| \le 2$\,kpc. After these selections, the final sample consists of approximately 55\,111 stars, with available abundance measurements for 55\,016 stars in C, 55\,042 in O, 55\,047 in Mg, 54\,765 in K, 55\,015 in Ca, 53\,445 in Ti, and 53\,267 in Cr.

\subsection{Hinkel et al. (2014) and Nissen et al. (2020) datasets}

In addition to the APOGEE data, we also consider abundance measurements from smaller, high-resolution spectroscopic surveys targeting Galactic disc stars in the optical wavelength regime (\citealt{hinkel2014,nissen2020}). These datasets provide an important complementary perspective, as they are based on observational setups with significantly higher spectral resolution than APOGEE ($R/R_{\rm APOGEE} > 10$), which may lead to systematic differences in the derived abundances (e.g. \citealt{spina2022,Hegedus22}).\\

The sample of \citet{hinkel2014} consists of 3058 thin disc stars in the solar neighbourhood. These stars belong to several surveys used to produce the Hypatia Catalog (see \citealt{hinkel2014} for a comprehensive list of literature sources adopted in Hypatia and the number of stars that matched the criteria and elemental abundances measured from each source), and are selected based on distance and spectral type. Namely, they select main sequence F/G/K/M-type stars within 150 pc of the Sun. The spectra were retrieved using a variety of instruments, with a resolving power spanning from R $\approx 30\,000$ to R $\approx 120\,000$, and a signal-to-noise of $\sim$100-200 per spectral pixel, in general. We refer to \citet{hinkel2014} for a complete discussion of the comparison of data sets. In this work, we adopt the published stellar abundances of Fe, O, Mg, Ca, Ti, K and Cr for the full sample. We should note that not all stars have the measured abundances for the elements analysed in this paper, instead there are 1\,709 abundances available for C, 2\,125 for O, 1\,604 for Mg, 284 for K, 1\,681 for Ca, 2\,237 for Ti and 1\,286 for Cr.

The abundance data from \citet{nissen2020} are derived from very high-resolution ($R > 100,000$) and very high signal-to-noise (S/N $> 600$) spectra of 72 solar twin stars, obtained with the HARPS and HARPS-N spectrographs at the La Silla 3.6,m and TNG telescope, respectively. These observations provide some of the most precise elemental abundance measurements currently available for stars in the solar neighbourhood. For our analysis, we adopt the abundances of Fe, C, O, Mg, Ca, Ti, and Cr reported for all stars in the \citet{nissen2020} sample.

\begin{table}[]
 \caption{List of models adopted in this work.}
    \centering
    \begin{tabular}{ccc}
    \hline
      Model & Stellar Yields & \% of massive binaries \\
    \hline
        R0  & R10  & -    \\
        F0  & FM23 & 0\%  \\
        F70 & FM23 & 70\% \\
        M0  & Ma25  & 0\%  \\
        M70 & Ma25  & 70\% \\
    \hline    
    \end{tabular}
    \label{tab_model}
\end{table}

\section{Results and discussion}
\label{res_sec}
In this section, we present the results of our chemical evolution models obtained by exploring different prescriptions for CC-SNe, including the newly implemented metallicity-dependent yields for massive binary stars. 
The model prescriptions adopted in this work are summarised in table~\ref{tab_model}. The first column lists the model names, while the second column specifies the massive star yield sets employed. The third column reports the fraction of massive binaries assumed when adopting the Ma25 or the FM23 yields. 

Throughout the paper, the model labelled with “R0” is referred to as Reference Model, as it adopts well-established yields for single massive stars from the literature, as described in section~\ref{subsec_yields} (see R10). The remaining models employ either the FM23 massive star yields (F0 and F70) or the Ma25 stellar yields as functions of metallicity (M0 and M70). For these models, the yields from other stellar sources are identical to those adopted in the R0 model (see section~\ref{subsec_yields}).

It is important to emphasise that the Reference Model (R0) shares the same physical assumptions as the other models described in section~\ref{mod_sec}, differing only in the adopted massive star yields, as detailed in section~\ref{subsec_yields}.
\\ 

\begin{figure}
    \centering
    \includegraphics[scale=0.35]{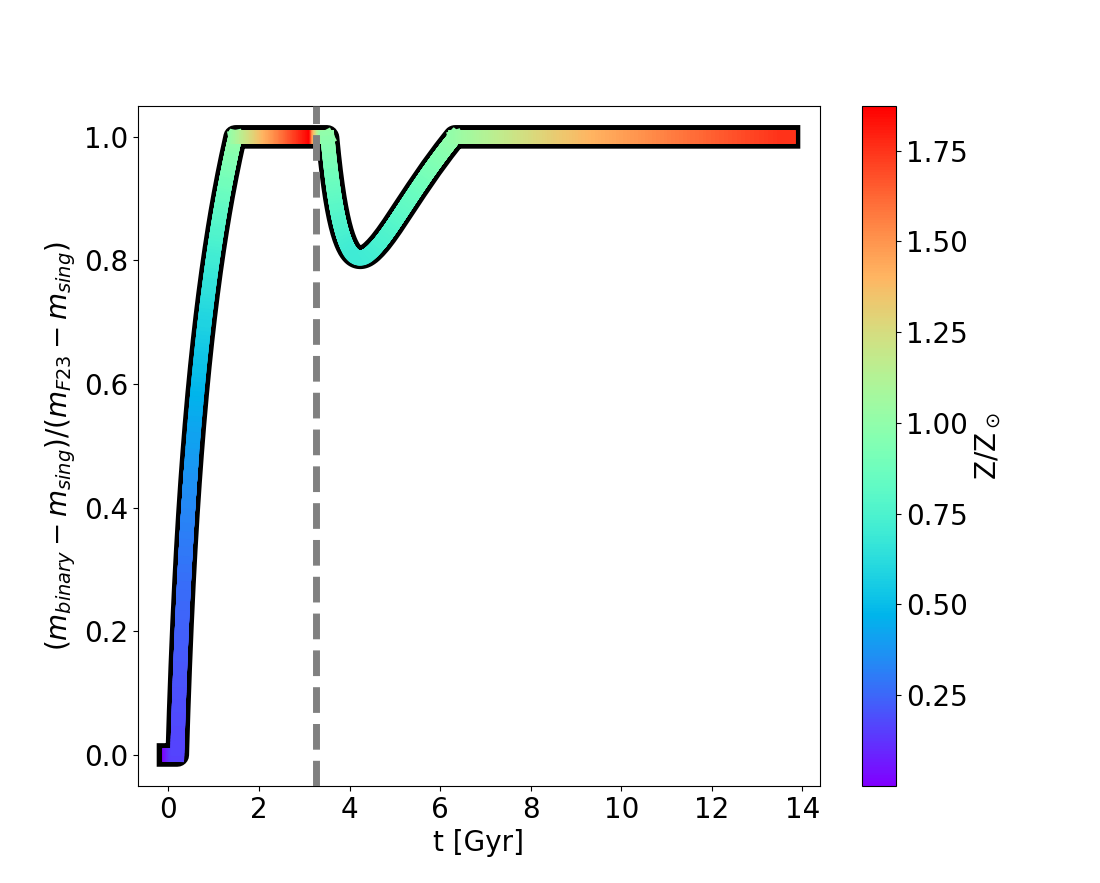}
    \caption{$(m_{binary}(Z)-m_{sing})/(m_{F23}-m_{sing})$ as a function of evolution time computed by model F70, colour-coded with metallicity. Thin gray, dashed line indicates the beginning of the second infall episode.}
    \label{fig_maeq}
\end{figure}

To better understand the difference between the yields adopted in this work and those of Paper~I, in fig.~\ref{fig_maeq} we show  the difference between FM23 yields for binary stars obtained from eq.~\ref{maeq} and their single star counterpart as a function of time as computed by our model F70, normalised to the difference between \citet{farmer2023} yields for both massive and binary stars. 
The line is colour coded with metallicity as computed by our F70 model. As we can see, at very low metallicity (early times), the yields of single and binary massive stars are the same. The difference between the two grows with time during the first infall event up until solar metallicity at around t$\sim$1.5 Gyr. When the second infall begins (t=3.25 Gyr), its effect is to dilute the gas and decrease its metallicity, causing the difference between FM23 binary yields and their single star counterpart to decrease. Shortly after star formation starts during the second infall event, gas metallicity begins to grow again as new massive stars enrich the ISM, and we reach solar metallicity again at t$\sim$6.5 Gyr.

\subsection{Comparison with physical Galactic quantities and solar abundances}
\label{pres_subsec}

\begin{figure*}
    \sidecaption
    \includegraphics[scale=0.39]{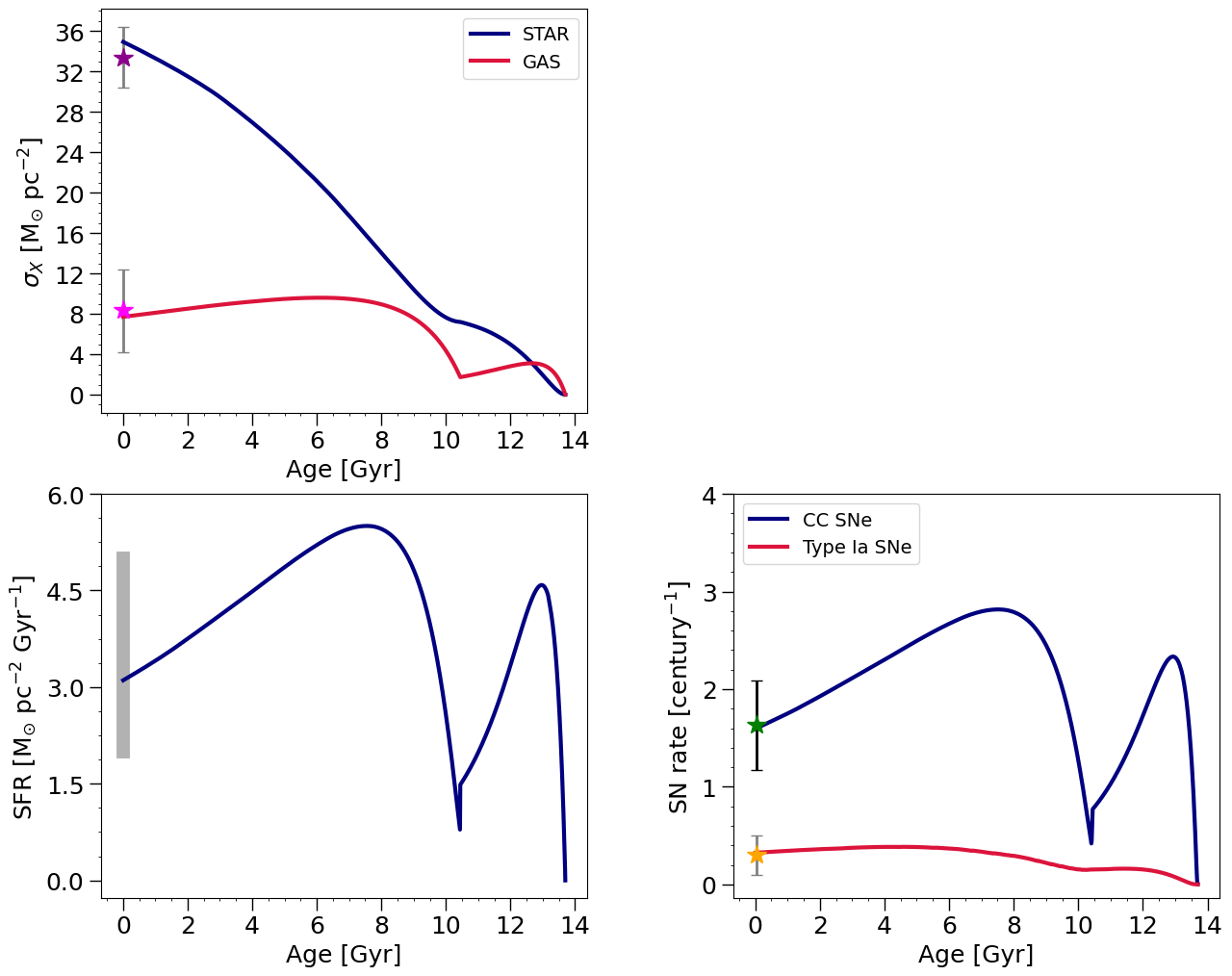}
     \caption{Galactic observables computed by our Reference Model (R0) compared with present-day data. Upper panel: evolution of the surface mass density of gas ($\sigma_{gas}$, red line) and stars ($\sigma_\star$, blue line). The purple star indicates the present day $\sigma_\star$ value as computed in \citet{mckee2015}, while the magenta star represents the present day $\sigma_{gas}$ value, obtained averaging between the \citet{dame1993} and \citet{Nakanishi2003,Nakanishi2006} for the solar vicinity as presented in \citet{palla2020}. The grey bar represents the 1 $\sigma$ error. Lower left panel: predicted time evolution of the SFR. The shaded grey area indicates the measured range in the solar neighbourhood as given by \citet{prantzos2018}. Lower right panel: evolution of SNe\,Ia (red line) and CC-SNe rates (blue line) predicted by our Reference Model (R0). The orange and green star represents the observed SNe\,Ia rate reported in \citet{cappellaro1997} and the observed CC-SNe rates from \citet{rozwadowska2021} respectively. We report in grey the 1 $\sigma$ error bars.}
    \label{fig_obs}
\end{figure*}

In  fig.~\ref{fig_obs}  we show a comparison between our Reference Model (R0) predictions for  Galactic observables such as the present day surface mass density of gas and stars, the SFR and the SN rates (II, Ia).
We emphasise that these predictions are independent of the adopted stellar yields and of the assumed fraction of massive binaries in the IMF. Consequently, they are identical for all models. 
In the upper panel, we present the temporal evolution of the surface mass density of gas and stars in the local thin disc. 
We recall that we impose a present day total surface mass density consistent with the value reported by \citet{mckee2015}, as described in section~\ref{mod_sec}. 
Our model R0 predicts a present-day stellar surface density of 34.9 M$_\odot\,\mathrm{pc}^{-2}$, well in agreement with the value of 33.4 $\pm$ 3 M$_\odot\,\mathrm{pc}^{-2}$ measured by \citet{mckee2015}, and again the model predicts a present-day gas surface density of 7.71 M$_\odot\,\mathrm{pc}^{-2}$, in agreement  with the observational estimates, as highlighted in the left panel of fig.~\ref{fig_obs}.

The lower-left panel of fig.~\ref{fig_obs} displays the temporal evolution of the SFR predicted by our model. The present-day SFR value of 3.11 M$_\odot\,\mathrm{pc}^{-2}\,\mathrm{Gyr}^{-1}$ lies within the range of 2–5 M$_\odot\,\mathrm{pc}^{-2}\,\mathrm{Gyr}^{-1}$ commonly adopted as a constraint for chemical evolution models of the solar neighbourhood \citep{matteucci2012, prantzos2018, spitoni2024}.

The lower-right panel of fig.~\ref{fig_obs} shows the time evolution of the SNe\,Ia and CC-SNe rates. For comparison, we adopt the present-day observational estimates for the solar vicinity of $1.63 \pm 0.46$ per 100 yr for CC-SNe \citep{rozwadowska2021} and $0.30 \pm 0.20$ per 100 yr for SNe\,Ia \citep{cappellaro1997}. 
Our model well reproduces both the observed SNe\,Ia rate and CC-SN rate, predicting present-day values of 0.32 per 100 yr and 1.59 per 100 yr respectively.\\ 

We continue by showing the model predictions obtained for the solar abundances, namely the ISM abundances at the time of the birth of our Sun. Model abundances are taken at an age of $4.5$ Gyr ago and are compared with measured solar abundances as obtained by \citet{asplund09}. 
Table~\ref{tab_abund} reports the solar abundances predicted by the two-infall model, expressed as abundances by number ($12+\log({\rm X}/{\rm H})$), for 22 chemical elements from He to Zn. As described in section~\ref{mod_sec}, the adopted chemical evolution framework has been extensively tested in previous studies and has been shown to successfully reproduce many observational constraints  such as the present-day star formation and SN rates, surface mass densities of stars and gas (see fig.~\ref{fig_obs}), abundance patterns and stellar ages in the solar neighbourhood (e.g. \citealt{spitoni2019, spitoni2020, palla2020, molero2023}). For this reason, the model predictions provide a meaningful basis for comparison with the observed solar abundances.
In table~\ref{tab_abund}, column~1 lists the chemical species, column~2 the observed solar abundances, while columns~3, 4, 5, 6 and~7 present the corresponding predictions from the different models of table~\ref{tab_model}.

Consistently with what already found in Paper~I, it is evident that, when adopting the FM23 yields, varying the assumed fraction of binary stars leads to negligible differences for most chemical elements. Noticeable variations are found only for $^{39}$K, $^{48}$Ti, and $^{51}$V.
In contrast, more significant differences emerge when comparing the results obtained using the FM23 yields with those derived from the yield of R10 (our Reference Model R0).

In particular, the Reference Model provides a better reproduction of the solar abundances of $^{19}$F, $^{23}$Na, $^{27}$Al, $^{28}$Si, $^{32}$S, $^{51}$V, $^{63}$Cu, and $^{64}$Zn. Conversely, the solar abundances of $^{12}$C, $^{24}$Mg, $^{39}$K, and $^{40}$Ca are more accurately reproduced by the models adopting the FM23 stellar yields. For $^{39}$K specifically, the FM23 yields reproduce the solar abundance satisfactorily only in the model assuming no contribution from binary stars (Model F0).

As for the predicted solar abundances of $^{12}$C when adopting the Ma25 yields for massive binary and single stars, we observe that both the M0 and M70 models (adopting no binary systems and 70\% binary systems in the IMF of massive stars, respectively), produce results that overestimate this element, with values well above 3 $\sigma$ relative to the observations.

\begin{figure*}
    \sidecaption
    \includegraphics[scale=0.266]{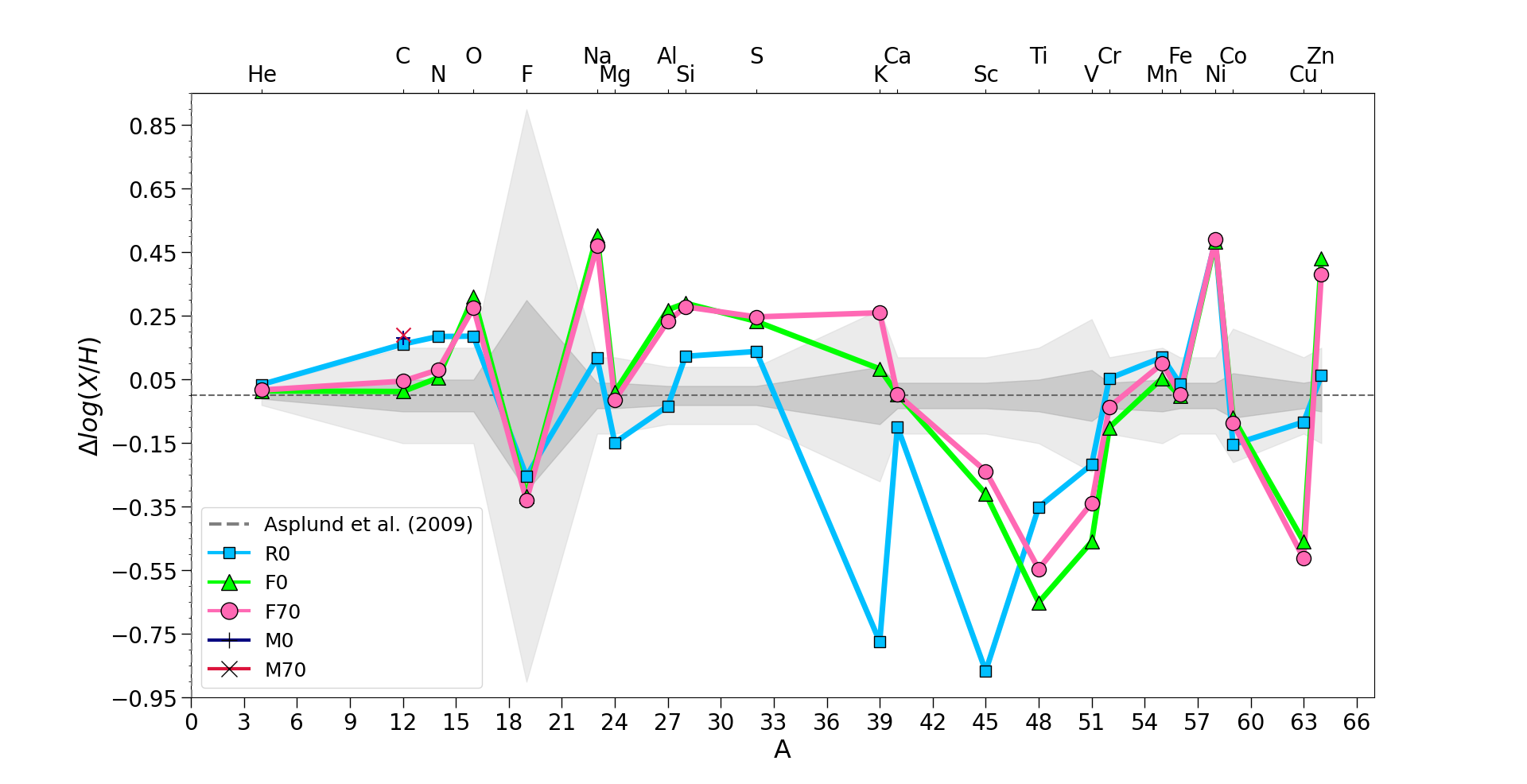}
    \caption{Differences in $\log$(X/H) abundance ratios for various elements predicted by our models, shown for different choices of stellar yields and adopted binary fractions (see legend). The thin gray dashed line marks the solar abundance ratios from \citet{asplund09}, while the gray and light gray shaded regions indicate the corresponding $\sigma$ and 3 $\sigma$ observational uncertainties respectively.}
    \label{fig_deltabun}
\end{figure*}

These discrepancies are further highlighted in fig.~\ref{fig_deltabun}, which shows the differences between the predictions of all our models and the solar abundances as reported by \citet{asplund09} for each element listed in table~\ref{tab_abund}. In this figure, the chemical elements are identified by the atomic mass number ($A$) of their dominant isotopes.

\begin{table}[]
\caption{Solar abundances as predicted by the two-infall model for the different yields tested in this work (see table~\ref{tab_model}).}
\label{tab_abund}
\resizebox{\columnwidth}{!}{
\begin{tabular}{|l|r|l|l|l|l|l|}
\hline
\multicolumn{1}{|c|}{\multirow{2}{*}{Element}} & \multicolumn{1}{c|}{\multirow{2}{*}{Asplund}} & \multicolumn{5}{c|}{Model}                                                                                                                      \\[0.03cm] \cline{3-7}
\multicolumn{1}{|c|}{}                          & \multicolumn{1}{c|}{}                         & \multicolumn{1}{r|}{R0}  & \multicolumn{1}{r|}{F0}  & \multicolumn{1}{r|}{F70} & \multicolumn{1}{r|}{M0}  & \multicolumn{1}{r|}{M70}\\[0.03cm] \hline

$^{4}$He   & 10.93 $\pm$ 0.01 & 10.96 & 10.94 & 10.95 & - & - \\ \hline
$^{12}$C   & 8.43  $\pm$ 0.05 & 8.59  & 8.44  & 8.48  & 8.61 & 8.62 \\ \hline
$^{14}$N   & 7.83  $\pm$ 0.05 & 8.01  & 7.88  & 7.91  & - & - \\ \hline
$^{16}$O   & 8.69  $\pm$ 0.05 & 8.88  & 9.00  & 8.96  & - & - \\ \hline
$^{19}$F   & 4.56  $\pm$ 0.30 & 4.31  & 4.24  & 4.23  & - & - \\ \hline
$^{23}$Na  & 6.24  $\pm$ 0.04 & 6.36  & 6.74  & 6.71  & - & - \\ \hline
$^{24}$Mg  & 7.60  $\pm$ 0.04 & 7.45  & 7.61  & 7.58  & - & - \\ \hline
$^{27}$Al  & 6.45  $\pm$ 0.03 & 6.42  & 6.72  & 6.68  & - & - \\ \hline
$^{28}$Si  & 7.51  $\pm$ 0.03 & 7.63  & 7.80  & 7.79  & - & - \\ \hline
$^{32}$S   & 7.12  $\pm$ 0.03 & 7.26  & 7.35  & 7.37  & - & - \\ \hline
$^{39}$K   & 5.03  $\pm$ 0.09 & 4.25  & 5.11  & 5.29  & - & - \\ \hline
$^{40}$Ca  & 6.34  $\pm$ 0.04 & 6.24  & 6.34  & 6.34  & - & - \\ \hline
$^{45}$Sc  & 3.15  $\pm$ 0.04 & 2.34  & 2.77  & 2.82  & - & - \\ \hline
$^{48}$Ti  & 4.95  $\pm$ 0.05 & 4.60  & 4.30  & 4.40  & - & - \\ \hline
$^{51}$V   & 3.93  $\pm$ 0.08 & 3.71  & 3.47  & 3.59  & - & - \\ \hline
$^{52}$Cr  & 5.64  $\pm$ 0.04 & 5.69  & 5.54  & 5.60  & - & - \\ \hline
$^{55}$Mn  & 5.43  $\pm$ 0.05 & 5.55  & 5.48  & 5.53  & - & - \\ \hline
$^{56}$Fe  & 7.50  $\pm$ 0.04 & 7.54  & 7.50  & 7.50  & - & - \\ \hline
$^{58}$Ni  & 6.22  $\pm$ 0.04 & 6.70  & 6.70  & 6.71  & - & - \\ \hline
$^{59}$Co  & 4.99  $\pm$ 0.07 & 4.84  & 4.92  & 4.90  & - & - \\ \hline
$^{63}$Cu  & 4.19  $\pm$ 0.04 & 4.11  & 3.73  & 3.67  & - & - \\ \hline
$^{64}$Zn  & 4.56  $\pm$ 0.05 & 4.62  & 4.99  & 4.94  & - & - \\ \hline
\end{tabular}
}
\end{table}

\subsection{Chemical abundance patterns}

In this section, we present the model predictions for the [X/Fe] versus [Fe/H] abundance ratios obtained by adopting different prescriptions for massive star yields.

In the following, we present the results for the chemical elements most relevant to this study. We focus on those elements for which extensive observational data are available and for which significant differences arise in the evolution of the [X/Fe] versus [Fe/H] relations when adopting different yield prescriptions. In particular, following Paper~I, we present the model results for C, O, Mg, Ca, Ti, K and Cr. Before looking at the [X/Fe] abundance diagrams, it is worth highlighting that in the following figures we consistently use the same colour scheme for the four models as adopted in fig.~\ref{fig_deltabun}.

It is also worth noting that the [X/Fe] versus [Fe/H] diagrams have to be interpreted according to the time-delay model (\citealt{tinsley1980, greggio1983, matteucci1986}; \citealt{matteucci2012,matteucci2021,matteucci2026}). The time-delay interpretation is grounded on the fact that the [Fe/H]-axis can be interpreted as a time evolution axis. Therefore,  at low metallicities (hence at earlier times), there is a predominant contribution to metals from massive stars and only at larger metallicities there is a substantial production of Fe from SNe\,Ia, which start exploding with a time delay relative to CC-SNe and can have explosion times of as long as a Hubble time. 
According to this, the [$\alpha$/Fe] ratios show the so-called plateau at low metallicities ([Fe/H] $\lesssim-1.0$ dex in the MW disc), and then decline for larger metallicities.
If the element considered is instead produced in larger amounts by SNe\,Ia and/or by LIMS, the change in slope at intermediate to high metallicity is less marked and becomes null or even positive when the element is produced exactly in the same proportions as Fe.
We note that the behaviour of the above elements relative to Fe in the two-infall evolutionary scenario, where
there is a natural gap in the SFR between the formation of the thick and thin discs, results in the loops observed in the  figures.
This behaviour is the consequence of a delayed second gas infall, which dilutes the ISM with primordial gas, lowering the [Fe/H] ratio and leaving the [X/Fe] unchanged. The metal abundance is then increased again thanks to the subsequent episode of star formation (see also \citealt{spitoni2019}).\\

\subsubsection{$\alpha$-elements}

Following the work of Paper~I, we begin by analysing the predicted abundance patterns for the most relevant $\alpha$-elements, namely $^{12}$C, $^{16}$O, $^{24}$Mg, $^{40}$Ca and $^{48}$Ti. We should note that all model outputs are normalised to \citet{asplund09} solar abundances, in agreement with the data presented in section~\ref{data_sec}.

\begin{figure}
    \centering
    \includegraphics[scale=0.40]{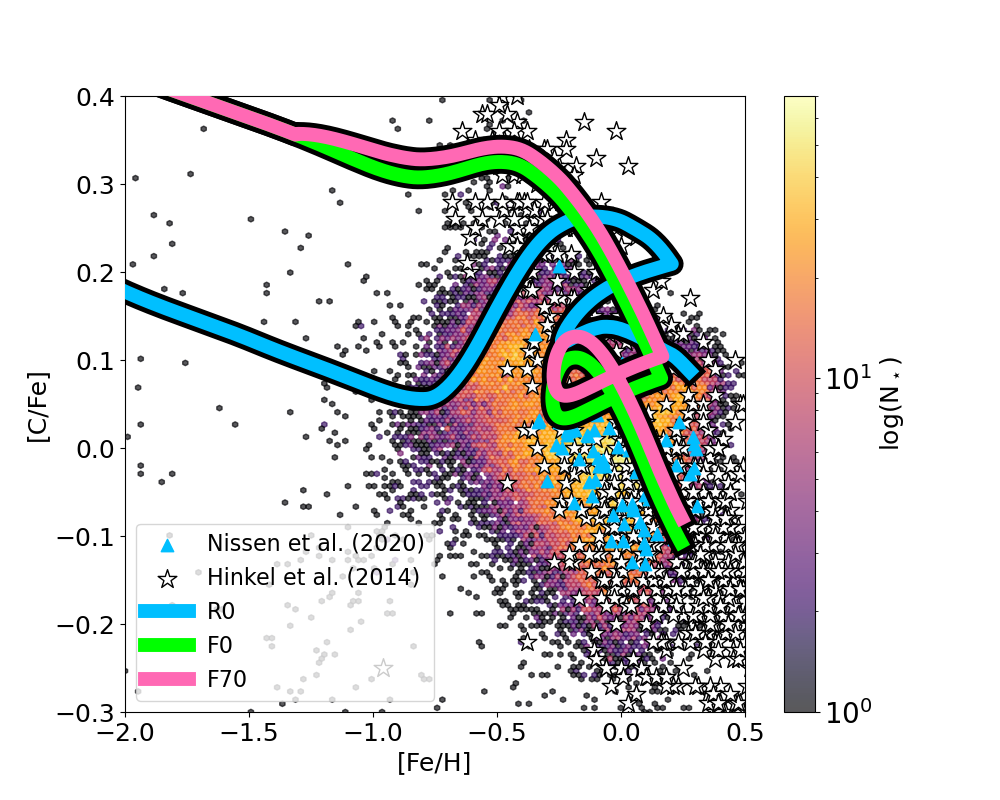}
    \caption{[C/Fe] vs [Fe/H] ratios for different stellar yields and percentage of binaries (see legend), and compared with data from  \citet[white stars]{hinkel2014}, \citet[azure triangles]{nissen2020} and APOGEE \citep{apogeedr172022}.
    }
    \label{fig_CFe}
\end{figure}

In fig.~\ref{fig_CFe}, we compare results for [C/Fe] versus [Fe/H] from our Reference Model (R0) and the F0 and F70 models which adopt the FM23 yields for different percentages of binaries in the IMF of massive stars (namely 0\% and 70\% respectively).
 
We compare our model results with data from stars in the solar vicinity, as described in section~\ref{data_sec}.
All the models displayed show some difficulty in reproducing the trends shown by the data.
In particular, the results of the Reference Model R0 predict a higher C abundance than the data from APOGEE (\citealt{apogeedr172022})  at [Fe/H] $>-0.5$ dex, while they agree with the observed trend at lower metallicities (see also \citealt{romano2010,romano2019}). The sudden increase in carbon abundance at higher metallicity is due to the contribution of low-mass AGB stars to the carbon enrichment (see also \citealt{romano2019,ventura22}), while it is almost hidden in models adopting the FM23 yields. 
On the other hand, as already observed in Paper~I, the models using FM23 yields, even with the new prescriptions, produce results largely overestimating the observed [C/Fe] ratio at low metallicities, although they are in relatively good agreement with the sample of APOGEE \citep{apogeedr172022} data for solar and super-solar metallicities. Therefore, the FM23 yields are better tracers of the C enrichment at high metallicities, whereas those of R10 are in better agreement with the trends observed at lower metallicities. When comparing our model results with high resolution surveys, we see that both F0 and F70 models adopting FM23 yields are able to reproduce the \citet{nissen2020} and the low-metallicity, high $\alpha$ stars from \citet{hinkel2014}, while their predictions do not reproduce the low-$\alpha$ component of the thin disc stars from \citet{hinkel2014}. The Reference Model predictions, on the other hand, have difficulties to reproduce the observed pattern in both data surveys.\\

\begin{figure}
    \centering
    \includegraphics[scale=0.40]{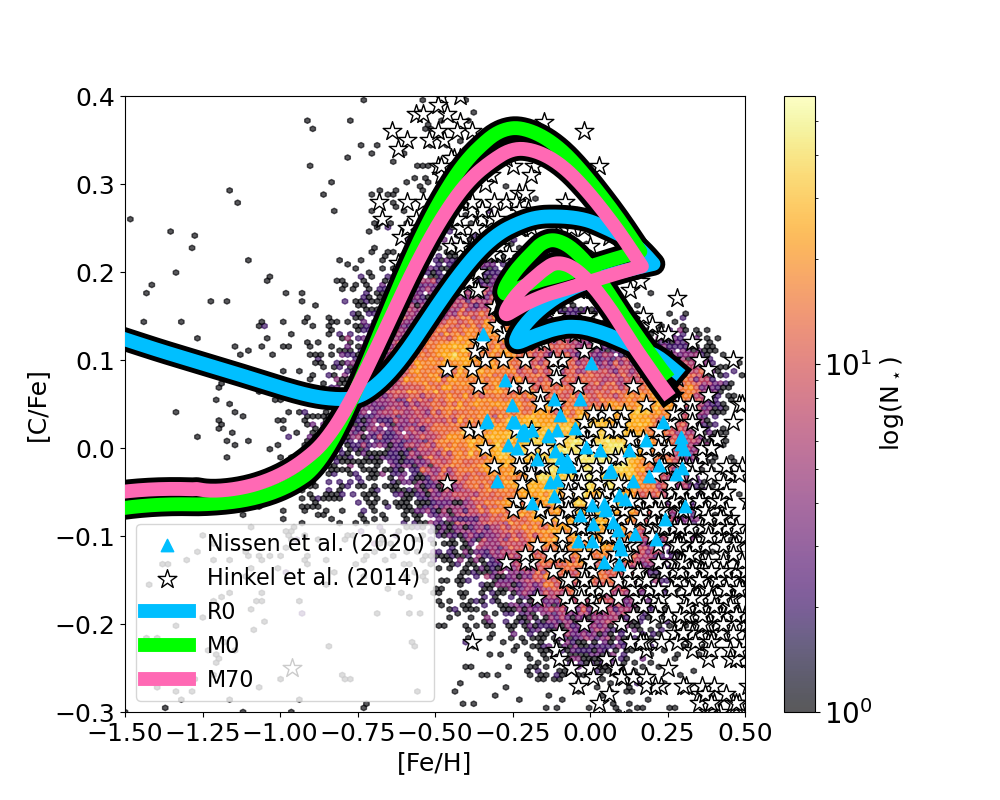}
    \caption{Same as fig.~\ref{fig_CFe} but for Ma25 yields. Data are from 
    \citet[white stars]{hinkel2014}, \citet[azure triangles]{nissen2020} and 
    APOGEE \citep[coloured according to their number density, see colourbar]{apogeedr172022}.}
    \label{fig_CFeMa}
\end{figure}

In fig.~\ref{fig_CFeMa}, we compare results for [C/Fe] versus [Fe/H] from our Reference Model (R0) and the M0 and M70 models that adopt the Ma25 stellar yields for different percentages of binaries in the IMF of massive stars (namely 0\% and 70\% respectively). Here the discrepancy between model predictions and observed abundance patterns becomes even more evident for all data surveys, with the model predictions from M0 and M70 only being able to reproduce the high-$\alpha$ pattern from \citet{hinkel2014}. While for the F0 and F70 models the bump in C production caused by low-mass AGB stars is largely hidden, when adopting the Ma25 yields, it becomes more evident than in the Reference Model. The Ma25 yields for $^{12}$C from massive binary stripped and single stars predict a sub-solar abundance of C at lower metallicity ([Fe/H] $\lesssim$ -0.8 dex), with most of the contribution at higher metallicity coming from low-mass AGB stars. Even without the limitation of the prescriptions adopted for FM23, the predicted abundance pattern for both M0 and M70 model are very similar, almost overlapping at all metallicities and diverging only after the beginning of the second infall event. None of the models is able to reproduce the observed abundances.\\

\begin{figure}
    \centering
    \includegraphics[scale=0.40]{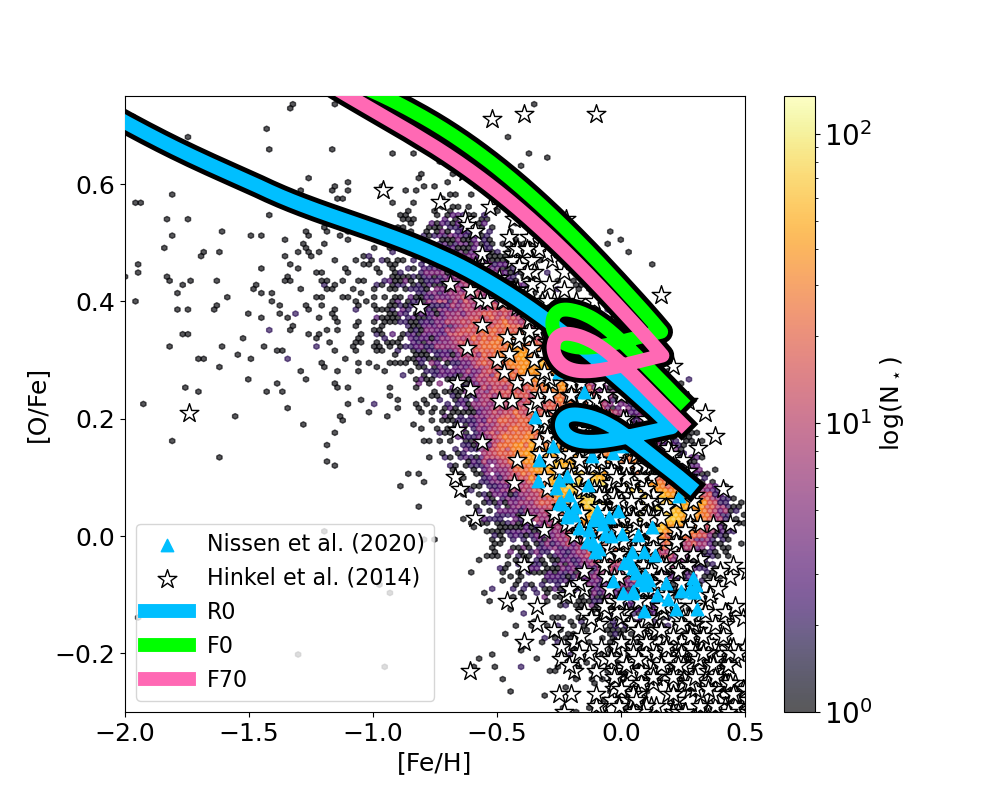}
        \caption{Same as fig.~\ref{fig_CFe} but for [O/Fe]. Data are from 
    \citet[white stars]{hinkel2014}, \citet[azure triangles]{nissen2020} and 
    APOGEE \citep[coloured according to their number density, see colourbar]{apogeedr172022}.}
    \label{fig_OFe}
\end{figure}

In fig.~\ref{fig_OFe}, we compare the model predictions for the [O/Fe] versus [Fe/H] relations.  
The model predictions generally overestimate the observed [O/Fe] trends in the various stellar samples. In particular, neither the predictions from Model F0 (adopting FM23 yields for single stars only) nor Model F70 (adopting FM23 yields and assuming a 70\% fraction of massive binaries) are able to reproduce any of the survey data. On the other hand, the predictions of the Reference Model R0 provide a better match to the observed trends at low metallicity, although they still overpredict the [O/Fe] ratio around solar metallicity.

We should note that models adopting the stellar yields of FM23 predict a higher abundance of O when adopting a lower fraction of binaries, in contrast to what occurs for carbon. This behaviour arises because an enhanced carbon production in massive stars leads to a reduced oxygen yield: a larger fraction of carbon is ejected through stellar winds (see \citealt{farmer2021, farmer2023}) and is therefore no longer processed into oxygen during later burning stages. For this same reason the Reference Model predicts a lower abundance of O, making it a better tracer for this element even if it still struggles to reproduce the observed abundances.  
These results suggest that the oxygen yields ejected by massive stars may be significantly overestimated in current stellar models, and that a lower oxygen production relative to iron may be required to reconcile chemical evolution models with the observed abundance patterns.\\

\begin{figure}
    \centering
    \includegraphics[scale=0.40]{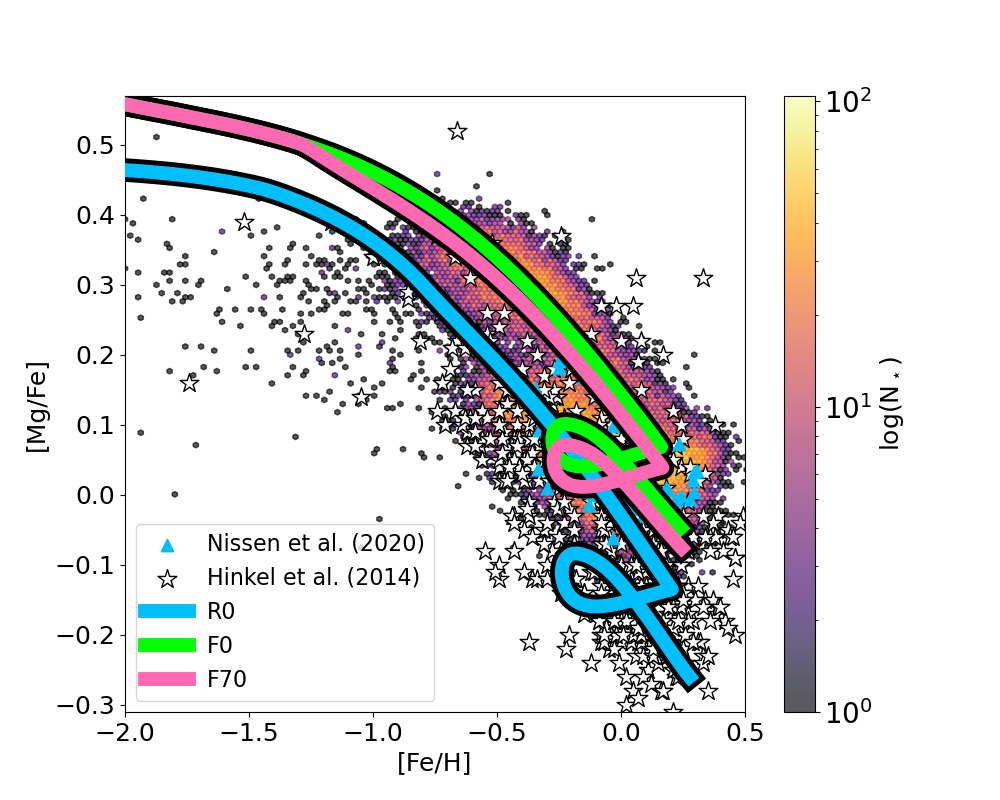}
    \caption{Same as fig.~\ref{fig_CFe} but for [Mg/Fe]. Data are from \citet[white stars]{hinkel2014}, \citet[azure triangles]{nissen2020} and 
    APOGEE \citep[coloured according to their number density, see colourbar]{apogeedr172022}.}
    \label{fig_MgFe}
\end{figure}

Fig.~\ref{fig_MgFe} presents the model predictions for $^{24}$Mg. The differences in magnesium yields between FM23 and R10 are primarily related to the inclusion of convective overshooting during carbon burning in \citet{farmer2023}, which is not considered in the \citet{koba2006} models (see also \citealt{Tominaga07}). This assumption has important consequences not only for carbon production—since the $^{12}$C yield is sensitive to the amount of carbon that survives previous burning stages—but also for the products of carbon burning, such as $^{20}$Ne, $^{23}$Na, and $^{24}$Mg, as well as for subsequent burning stages leading to heavier elements (see, e.g., \citealt{farmer2021, farmer2023}). These effects propagate into the chemical evolution models adopting the FM23 yields, resulting in systematically higher predicted [Mg/Fe] ratios relative to the Reference Model.

In particular, the predicted abundance pattern of the Reference Model (R0) displays a clear mismatch with both the \citet{nissen2020} and the APOGEE data \citep{apogeedr172022}, underestimating the observed [Mg/Fe] trend, especially at high metallicity (see also \citealt{palla2022} for a comparison with an independent observational sample). This is a well known problem, since in the current literature the  $^{24}$Mg yields from massive stars are generally too small to reproduce the observations (see \citealt{matteucci2021}). On the other hand, \citet{hinkel2014} data are  very well reproduced.

In contrast, the models adopting the FM23 yields provide a substantially better match to the data.

Notably, both the F0 model, which assumes no contribution from binary-stripped stars, and the F70 model, which assumes a 70\% fraction of massive binaries, reproduce remarkably well both the high-$\alpha$ and low-$\alpha$ sequences observed in the Galactic disc from both \citet{nissen2020} and APOGEE \citep{apogeedr172022} data sets, while both their predicted abundances struggle to reproduce most of \citet{hinkel2014} dataset, predicting an overall higher [Mg/Fe] relation throughout the whole metallicity range.

\begin{figure}
    \centering
    \includegraphics[scale=0.40]{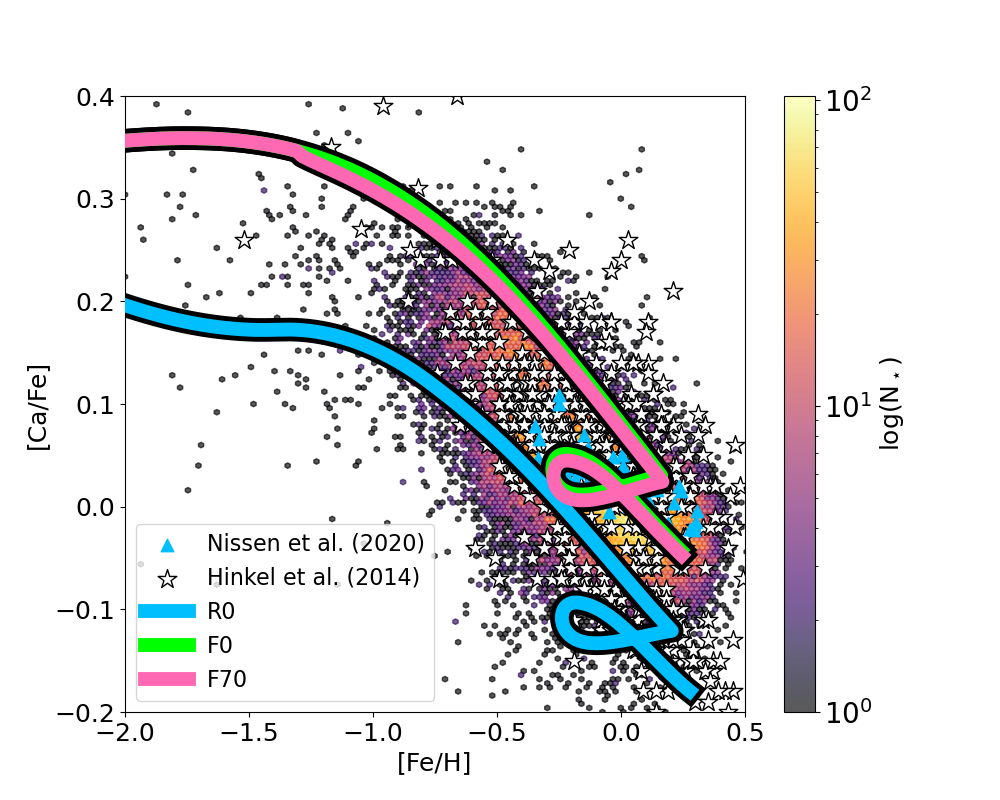}
    \caption{Same as fig.~\ref{fig_CFe} but for [Ca/Fe]. Data are from \citet[white stars]{hinkel2014}, \citet[azure triangles]{nissen2020} and 
    APOGEE \citep[coloured according to their number density, see colourbar]{apogeedr172022}.}
    \label{fig_CaFe}
\end{figure}

Another $\alpha$-element that shows quite significant variations in response to the different studied yields is Ca. In fig.~\ref{fig_CaFe}, we compare results from our models for [Ca/Fe] versus [Fe/H]. 
The model implementing R10 yields shows a lower [Ca/Fe] enrichment, with subsolar [Ca/Fe] values at high metallicities and an offset of the order of $\sim 0.15$ dex relative to FM23 yield sets.
When observing F0 and F70 model, we see that their predictions almost overlap for the whole [Fe/H] range, showing that binary stars produce the same contribution as their single star counterpart to the enrichment of Ca into the ISM.

When comparing our model results with data, we see that the Reference Model predictions do not agree with the observational data, as they clearly under-produce [Ca/Fe] at all metallicities (see also \citealt{romano2010}), except for the low-$\alpha$ component from \citet{hinkel2014}.
Conversely, the models adopting FM23 yields reproduce all the adopted data surveys very well, with no visible difference between the models at all metallicities, despite the variation in the percentage of binaries inside the chemical evolution model.\\

\begin{figure}
    \centering
    \includegraphics[scale=0.40]{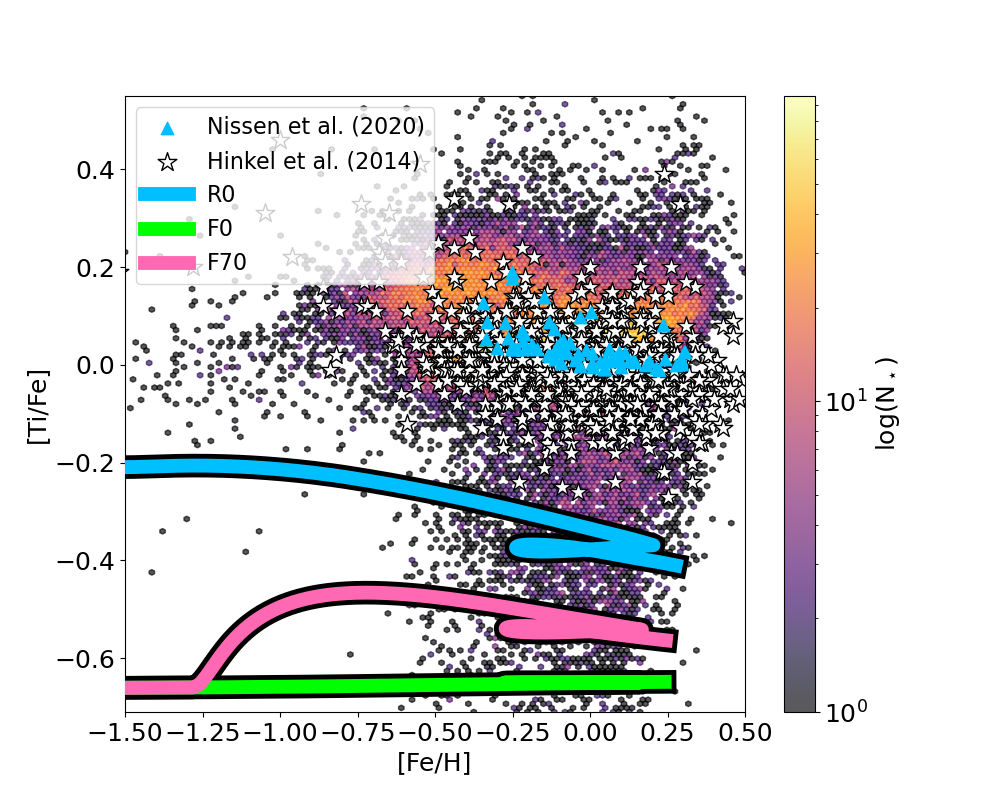}
    \caption{Same as fig.~\ref{fig_CFe} but for [Ti/Fe]. Data are from \citet[white stars]{hinkel2014}, \citet[azure triangles]{nissen2020} and 
    APOGEE \citep[coloured according to their number density, see colourbar]{apogeedr172022}.}
    \label{fig_TiFe}
\end{figure}

The final $\alpha$-element considered in this study is titanium, although its nucleosynthetic origin remains uncertain, as highlighted by several authors (see, e.g. \citealt{romano2010, prantzos2018, kobayashi2020}). It is worth emphasising that titanium production is highly sensitive to the assumption of spherical symmetry commonly adopted in one-dimensional CC-SN models. This limitation can only be addressed through the use of multidimensional simulations \citep{rauscher2002, magkotsios2010, harris2017, sandoval2021}.

Despite the differences among the adopted yield sets shown in fig.~\ref{fig_TiFe}, none of the models explored in this work are able to reproduce the observed stellar titanium abundances. The largest discrepancy is found for Model F0, which adopts FM23 yields for single massive stars. As discussed above, this disagreement may be alleviated by employing multidimensional CC-SN models. However, such an approach lies beyond the scope of the present paper.

\subsubsection{Potassium}

Turning now to the odd-Z elements, we focus on potassium, whose theoretical yields have long been known to significantly underestimate the observed Galactic abundance patterns (\citealt{romano2010, kobayashi2020} and references therein), even when additional mechanisms such as stellar rotation are taken into account (\citealt{prantzos2018}). Only the recent study by Paper~I has succeeded in reproducing the observed [K/Fe] abundance ratios without the need to invoke ad hoc assumptions. 

\begin{figure}
    \centering
    \includegraphics[scale=0.40]{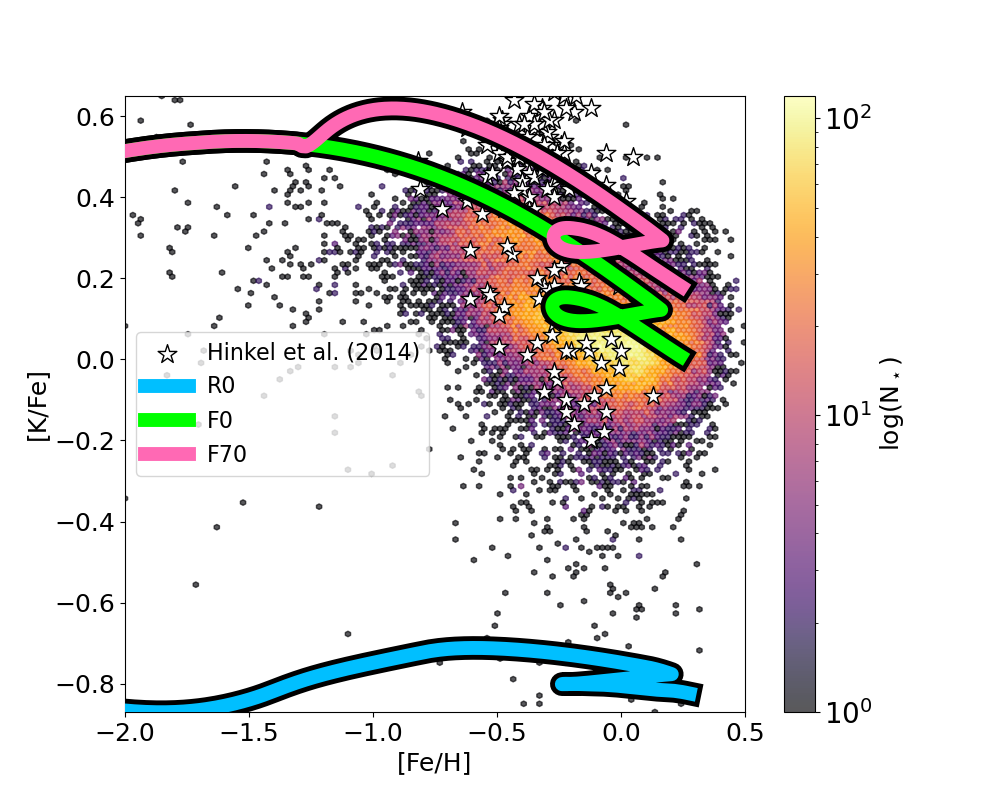}
    \caption{Same as fig.~\ref{fig_CFe} but for [K/Fe]. Data are from \citet[white stars]{hinkel2014} and APOGEE \citep[coloured according to their number density, see colourbar]{apogeedr172022}.}
    \label{fig_KFe}
\end{figure}

In fig.~\ref{fig_KFe}, we compare our model predictions for [K/Fe] versus [Fe/H], testing the newly proposed FM23 yields.
The differences between the yield sets arise because, in the FM23 yield grids, very high-mass stars in binaries produce lower amounts of K, whereas lower mass massive stars, which dominate the enrichment at higher metallicities, experience more favourable conditions for K production compared to single stars in the same mass range. As suggested by \citet{farmer2023}, potassium production is therefore strongly affected by the presence of binaries, with low-mass binary-stripped stars during their pre-supernova evolution being the main contributors.

The radically different behaviour between the Reference Model (R0), representative of the predictions obtained with the most widely used yields in the literature (\citealt{WW1995, koba2006, kobayashi2011, limongi2018}), and the models adopting FM23 yields has important implications for comparisons with the observed MW abundance patterns. 
While the Reference Model R0 severely underestimates the observed K abundances, Model F0 (adopting FM23 yields assuming only single stars) reproduces the observed [K/Fe] trend in the solar vicinity remarkably well, in agreement with the results of Paper~I, using solar metallicity yields by \citet{farmer2023}. Model F70, by contrast, slightly overestimates the observed [K/Fe] ratios and fails to reproduce the bulk of the data.

\subsubsection{Chromium}

For what concerns the Fe-peak group, the element showing the most significant variations between different models is Cr. 

\begin{figure}
    \centering
    \includegraphics[scale=0.40]{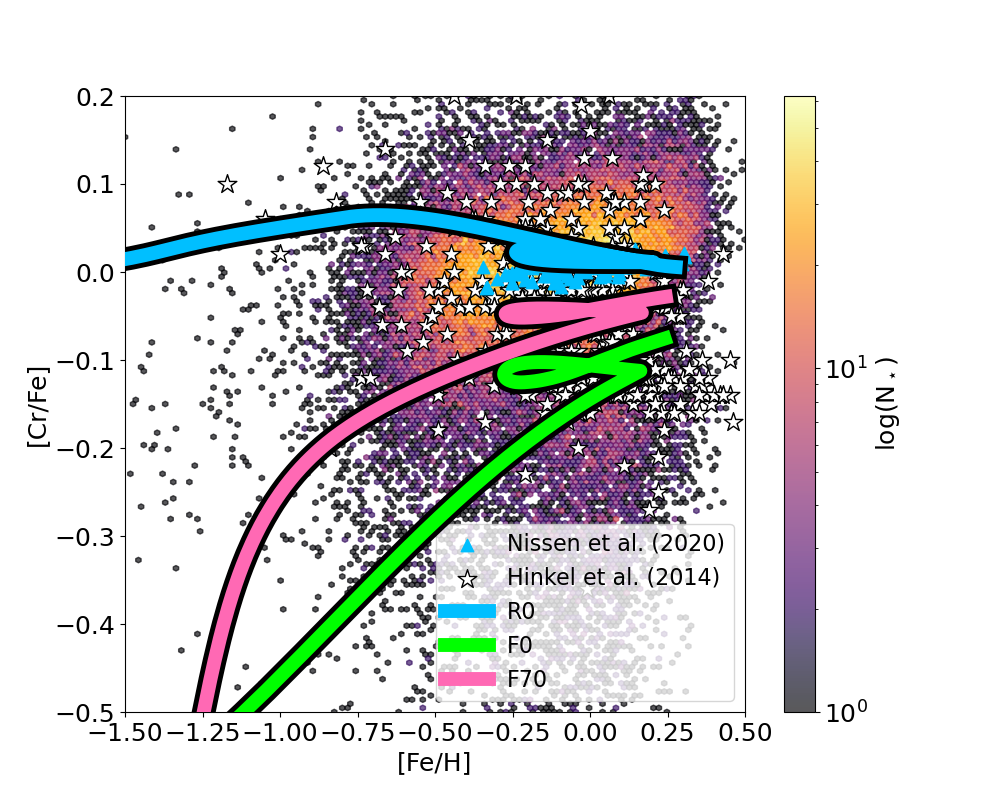}
    \caption{Same as fig.~\ref{fig_CFe} but for [Cr/Fe]. Data are from \citet[white stars]{hinkel2014}, \citet[azure triangles]{nissen2020} and      
    APOGEE \citep[coloured according to their number density, see colourbar]{apogeedr172022}.}
    \label{fig_CrFe}
\end{figure}

In fig.~\ref{fig_CrFe}, we can observe that the Reference Model R0 predicts an almost flat [Cr/Fe] pattern, with values remaining close to solar over the whole metallicity range. In contrast, models adopting the FM23 yields show markedly lower [Cr/Fe] ratios at low metallicity, followed by a gradual increase towards higher metallicities and approaching [Cr/Fe] $\sim -0.15$ dex at [Fe/H] $\gtrsim 0$. Model F70, assuming a 70\% fraction of massive binaries, exhibits a short and steep rise in [Cr/Fe], increasing from $\sim -0.5$ dex to $\sim -0.2$ dex between [Fe/H] $\sim -1.25$ and $\sim -1.0$ dex, and then continuing with a more moderate increase towards approximately solar values. As also discussed in Paper~I, this behaviour suggests that, within the FM23 stellar models, single massive stars contribute only marginally to Cr production relative to Fe. This is in contrast with the case of binary-stripped stars, which produce Cr in amounts comparable to Fe, in better agreement with most commonly adopted yield prescriptions in the literature (see also \citealt{prantzos2018, koba2020b, palla2021}).

When comparing our model predictions with data, the nearly flat trends predicted by the Reference Model and by the model adopting FM23 yields with the highest binary fraction are broadly consistent with the observational data for solar-neighbourhood stars described in section~\ref{data_sec}. This comparison clearly disfavours Model F0 as a suitable tracer of chromium evolution in the solar vicinity, as it predicts an increasing [Cr/Fe] trend that is not supported by APOGEE data. The remaining F70 Model provides a better qualitative match to the observed behaviour, although its predicted chemical tracks tend to slightly underestimate the overall Galactic Cr abundances.

\subsection{Discussion}

In this study, we have investigated the impact on the chemical evolution of the MW of new $^{12}$C yields from massive single and binary-stripped stars as computed by \citet{ma2025}, as well as the effects of metallicity-dependent corrections for the yields of other chemical elements as introduced in the same work, starting from the previously computed  \citet{farmer2023} yields.
In doing so, we have extended the analysis of Paper~I, where we tested the yields of \citep{farmer2023}  both for single and binary massive stars but  only for the solar chemical composition.
Earlier studies have investigated the role of interacting massive binaries in Galactic chemical enrichment (e.g., \citealt{dedonder&vanbeveren2002,dedonder2004}). These works generally concluded that the inclusion of massive binaries does not lead to dramatic changes in the predicted abundance evolution, with differences typically remaining within a factor of two compared to models neglecting binaries. In particular, Paper~I showed that variations in the adopted yields significantly affect both the predicted solar abundances and abundance patterns, whereas binarity itself introduces only modest changes for most of the chemical elements.
Building on the results of \citet{farmer2021} and \citet{farmer2023}, \citet{ma2025} investigated the metallicity dependence of carbon production in massive binary systems, finding that at low metallicity binary-stripped stars produce amounts of carbon comparable to those of their single-star counterparts. On the basis of this result, they proposed the prescription given in eq.~\ref{maeq} to reproduce this behaviour while still adopting the \citet{farmer2023} yield grids.

In the present analysis, we have shown that adopting yields from massive stars in binaries, rather than standard nucleosynthesis prescriptions for single massive stars, can lead to different predicted solar abundances and, for some elements, can improve the agreement with observations. However, we have also found that the metallicity-dependent prescriptions, introduced by \citet{ma2025}, significantly reduce the differences in the results predicted by models assuming different fractions of interacting massive binaries in the IMF. While these prescriptions address one of the main limitations of Paper~I, they remain an approximation, and therefore a fully consistent comparison with other yields available in the literature is not yet possible.
This limitation is particularly relevant for elements with a strong secondary component, such as $^{14}$N, the odd-$Z$ elements $^{27}$Al and $^{23}$Na, and $^{63}$Cu. 

A further limitation shared by both Paper~I and this work concerns the treatment of secondary stars in binary systems, which are assumed to evolve as single stars. This simplification originates from the yield calculations of \citet{farmer2023} and \citet{ma2025}, which do not provide nucleosynthetic yields for these binary products, instead treating it as a point mass that evolves alongside the primary up until the end of core He-burning. Once core He-burning ends, the authors do not consider the companion star and assume that no further interaction happens until core collapse. Nevertheless, secondary stars may contribute significantly to Galactic chemical enrichment if mass accretion allows stars initially below the core-collapse threshold ($M \lesssim 8\,M_{\odot}$) to explode as supernovae (e.g., \citealt{2004podsiadlowski}; \citealt[\citeyear{2019Zapartas}]{2017Zapartas}). Moreover, massive binary systems at low metallicity can be a source of primary $^{14}$N, because during the SN explosion the primary star ejects C-rich material, produced during its evolution, which is then accreted onto the secondary star and  then turned into primary $^{14}$N \citep{dedonder2004}. Therefore, ignoring the evolution of the secondary star may have a strong impact on the evolution of secondary elements such as $^{14}$N at low metallicity.
The treatment of binary evolution and the evolution of the secondary star in particular are also essential for the study of r-process elements, whose enrichment has a strong contribution from binary neutron star systems. In fact, merging neutron stars can produce substantial quantities of r-process elements, such as Eu, as shown by  \citet{dedonder2004, matteucci2014, simonetti2019, molero2021, palla2025}.
However, the treatment of binary companions is further complicated by uncertainties related to (i) the fraction of accreted mass that remains bound to the secondary star and (ii) the degree to which this accreted material is subsequently processed through nuclear burning (see, e.g., \citealt{2021Deckers}). 

Finally, an additional source of uncertainty arises from the choice of stellar evolution code used to compute the yields. Different codes adopt different treatments of key physical processes, such as convection and the mechanism driving supernova explosions, leading to non-negligible discrepancies in the predicted nucleosynthetic outputs.

For these reasons, we do not aim to draw firm conclusions regarding the quantitative role of massive binaries in Galactic chemical enrichment, nor to constrain the true fraction of massive binaries. This is further hindered by the lack of well-sampled grids of binary system parameters, such as mass ratios and orbital periods, and by the simplifying assumption that all massive binaries undergo Case~B mass transfer, producing binary-stripped primaries (see \citealt{Moe17} for a review of binary parameter distributions).

Instead, the primary goal of this work is to demonstrate that the inclusion of yields from massive stars in binary systems can influence the level of agreement between chemical evolution models and observed abundance patterns. In this sense, our study is intended both to motivate further developments in the modelling of stellar yields from binary systems, particularly through expanded grids in metallicity and binary parameters, and to provide a foundation for future investigations into the role of binary evolution in Galactic chemical evolution.

\section{Conclusion}
\label{conc_sec}

In this paper, we have investigated the impact of metallicity-dependent massive binary star yields within the framework of a well-tested chemical evolution model for the Milky Way \citep[see, e.g.,][]{spitoni2019, palla2020}. In particular, we have extended the analysis of Paper~I by adopting the prescriptions introduced by \citet{ma2025}, which allow the massive binary star yields computed by \citet{farmer2023} to be treated as a function of metallicity. We explored different assumptions for the fraction of massive binaries in the IMF, namely 0\% and 70\%, motivated by the results of Paper~I, which showed that variations in the massive binary fraction generally lead to only minor changes once binaries are included. We also tested the new $^{12}$C yields computed by \citet{ma2025} for massive binary-stripped and single stars as a function of metallicity. 

For comparison purposes, we also considered a Reference Model adopting metallicity-dependent yields for single massive stars taken from the literature \citep[][their model 15]{romano2010}, which have been widely used in previous chemical evolution studies. Our primary aim was to address one of the main limitations of Paper~I, namely the lack of metallicity dependence in the adopted massive star yields, and to assess the effects of the new prescriptions proposed by \citet{ma2025} on the chemical evolution of the solar vicinity.

Our main results can be summarised as follows:

\begin{itemize}

    \item models adopting the Ma25 yields tend to overproduce the solar carbon abundance and are unable to reproduce the observed [C/Fe] abundance pattern in the solar neighbourhood. Varying the fraction of massive binaries in the IMF does not significantly affect the predicted carbon abundances;

    \item consistently with the findings of Paper~I, when adopting the FM23 stellar yields, the differences arising from varying the massive binary fraction are negligible for most of the chemical elements considered. The adoption of the metallicity-dependent prescriptions of \citet{ma2025} further reduces the already small differences associated with different binary fractions, particularly at low metallicity;

    \item more substantial differences are observed when comparing the Reference Model (R0), which adopts already tested metallicity-dependent yields for single massive stars, with models employing FM23 yields for both single and binary stars. These differences affect both the predicted solar abundances and the overall abundance patterns;

    \item the solar abundances predicted by Model F0, which adopts FM23 yields, show improved agreement with observations for $^{4}$He, $^{12}$C, $^{14}$N, $^{24}$Mg, $^{39}$K, $^{40}$Ca, $^{55}$Mn, $^{56}$Fe, and $^{59}$Co. When assuming a 70\% fraction of massive binaries (Model F70), better agreement is found for $^{24}$Mg, $^{40}$Ca, $^{52}$Cr, and $^{56}$Fe. We note that both Model F0 and Model F70 predict identical solar abundances for $^{40}$Ca and $^{56}$Fe.

    \item Concerning the [X/Fe] versus [Fe/H] abundance patterns, all models struggle to reproduce both the [C/Fe] and [Ti/Fe] vs [Fe/H]. trends. In contrast, the [Mg/Fe] ratio is well reproduced by both Model F0 and Model F70. Similarly, the [Ca/Fe] versus [Fe/H] relation predicted by models adopting FM23 yields show very good agreement with the observational data, particularly at high metallicity. On the other hand, the [O/Fe] and [Cr/Fe] abundance ratios are best reproduced by the Reference Model, while Model F0 and Model F70 respectively overestimate and underestimate the observed trends.

    \item Model F0 successfully reproduces the observed $^{39}$K abundance trend without invoking ad hoc assumptions on the stellar yields, which are commonly required by most available yield prescriptions (e.g., \citealt{kobayashi2020}), confirming the results of Paper~I.

    \item For the remaining chemical elements, we do not find significant differences in either the predicted solar abundances or the [X/Fe] versus [Fe/H] relations when comparing models that adopt different stellar yield prescriptions.

\end{itemize}

\begin{acknowledgements}
F. Matteucci thanks I.N.A.F. for the 1.05.12.06.05 Theory Grant - Galactic archaeology with radioactive and stable nuclei.    F. Matteucci thanks also support from Project PRIN MUR 2022 (code 2022ARWP9C) “Early Formation and Evolution of Bulge and HalO (EFEBHO)” (PI: M. Marconi).
E. Spitoni and F. Matteucci thank I.N.A.F. for the  
1.05.24.07.02 Mini Grant - LEGARE "Linking the chemical Evolution of Galactic discs AcRoss diversE scales: from the thin disc to the nuclear stellar disc" (PI E. Spitoni).
E. Spitoni  thanks I.N.A.F. for the  1.05.23.01.09 Large Grant - Beyond metallicity: Exploiting the full POtential of CHemical elements (EPOCH) (ref. Laura Magrini).
M. Palla acknowledges support from HORIZON-INFRA-2024-DEV-01-01 – Research Infrastructure Concept Development, through the project WST: The Wide-Field Spectroscopic Telescope (Grant No. 101183153).
In this work, we have made use of SDSS-IV APOGEE-2 DR17 data. Funding for the Sloan Digital Sky Survey IV has been provided by the Alfred P. Sloan Foundation, the U.S. Department of Energy Office of Science, and the Participating Institutions. SDSS-IV acknowledges support and resources from the Center for High-Performance Computing at the University of Utah. The SDSS web site is  \href{www.sdss.org}{www.sdss.org}.
SDSS is managed by the Astrophysical Research Consortium for the Participating Institutions of the SDSS Collaboration which are listed at \href{https://www.sdss.org/collaboration/affiliations/}{www.sdss.org/collaboration/affiliations.} 
Finally, we thank the referee, Dany Vanbeveren, for his careful reading of the manuscript and useful suggestions that improved the paper.
 \end{acknowledgements}

\bibliographystyle{aa} 
\bibliography{pepe}

\end{document}